\documentstyle[twocolumn,epsf,aps]{revtex}
\draft
\begin{document}

\title{Entropy derivation for cluster methods
in non-Bravais lattices}
\author{Gy\"orgy Szab\'o}
\address
{Research Institute for Materials Science, H-1525 Budapest,
POB 49, Hungary}
\address{\em \today}
\address{
\centering{
\medskip \em
\begin{minipage}{15.4cm}
{}~~~The derivation of entropy for cluster methods is
reformulated by constructing the probability of a given
particle (spin) configuration as a self-consistent product
of cluster configuration probabilities. This approach gives
an insight into the nature of underlying approximations
involved at different levels of the cluster-variation
method. The graphical representation of the product allows
us to extend this method for non-Bravais lattices as it is
demonstrated on interstitial sites of body-centered- and
face-centered-cubic crystals.
%
\end{minipage}
}}
\maketitle
\narrowtext

\section{INTRODUCTION}
\label{sec:intro}

The cluster-variation method (CVM) is considered as a
generalization of the traditional mean-field
(Bragg-Wiliams\cite{bw}) theory taking only one-point
configurations into account. The first-neighbor pair
correlations were handled in the method developed by Bethe
for square lattice \cite{bethe}. The accuracy of this
technique were improved by Kramers and Wannier, who chose a
square cluster on which the probabilities of each
configuration were taken into consideration \cite{kw}. The
variational method was generalized by Kikuchi for different
lattice structure using a large class of clusters
\cite{kiku}. The above authors used combinatorial methods to
determine the entropy as a function of cluster configuration
probabilities.
Since then the CVM has been reformulated by several authors.
These formulations were motivated by the cumbersome
calculations when determining ensemble configurations for
large clusters in three-dimensional lattices. Barker showed
the equivalence between a generalization of quasi-chemical
approximation and CVM \cite{barker}. Hijmans and De Boer
developed a systematic scheme for obtaining the free energy
expressed by cluster variables \cite{hij}. Woodbury found
that many of the previous results were reproducible in a
direct manner using the composition law of information
theory \cite{wood}. Introducing a generalized cumulant
expansion of the entropy the CVM was reformulated by Morita
\cite{morita}. This latter formulation is strongly related
to the M\"obius inversion \cite{schl1,an,mm}.
A systematic counting and classification of clusters were
carried out by Gratias {\it et al.} \cite{grat} for general
crystal structures including some non-Bravais lattices.
Sanchez {\it et al.} developed a general formalism based on
the description of configurational cluster functions in
terms of an orthogonal basis in the multidimensional space
of site variables \cite{sdg}. Several authors investigated
the convergence of CVM when using larger and larger clusters
\cite{schl1,kkw,kb,schl2,peli}.

Recently the CVM is widely used to determine phase diagrams
in different systems (for reviews see the works by de
Fontaine \cite{cluster,cvm} and further references therein).
In the present work we suggest a simple way for the entropy
derivation which is not related to the Gibbs formalism and
may be used for dynamical (non-equilibrium) systems too.
Furthermore this method is applicable to non-Bravais lattices
which are very important for practice (e.g. superionic conductors,
metal-hydrogen systems, intercalation alloys, etc). For this
purpose the CVM is reformulated by showing how to construct
the probability of a given particle (spin) configuration as
a product of cluster configuration probabilities taking the
translation symmetries and self-consistency into account.
These conditions can be satisfied by using different
approximations equivalent to those mentioned above. Now we
concentrate on the entropy per elementary cells which is
related to the conditional entropy introduced in information
theory \cite{sw}.
As a result the entropy as well as the energy are expressed
as a function of the probabilities of cluster
configurations. Thus we have a free energy to be minimized
with respect to these variables for obtaining thermodynamic
equation of states.

In the subsequent section the essence of this method is
illustrated on the one-dimensional lattice. This idea is
adapted for the two-dimensional system in
Sec.~\ref{sec:square}. Here, a graphical representation of
the construction of configuration probabilities will be
introduced. This technique is used to obtain entropy
expressions for two non-Bravais lattices formed by the
interstitial sites in body-centered- and face-centered-cubic
lattices in Secs. \ref{sec:BCC} and \ref{sec:FCC}. Finally
the results are summarized in Sec.~\ref{sec:conc}.

\section{One-dimensional lattice}
\label{sec:1d}

In the one-dimensional model  a site variable $\eta_i$
($i=1,...,N$) denotes the state of the $i$th lattice point.
In the lattice-gas formalism ($\eta_i=0$ or 1), however, the
subsequent formulae remain valid for those systems
characterized by $Q$ states at each site (e.g. $\eta_i=0$,
$1$, $Q-1$). The function $f_N=f(\eta_1,...,\eta_N)$
describes the probability of a configuration specified by
the $\eta_i$ variables. Similarly, we introduce a cluster of
$k$ subsequent sites on which the probability of a
configuration is defined by $p_k(\eta_1, \ldots , \eta_k)$.
If the system is translation invariant, then
\begin{equation}
\sum_{\eta_i \atop i<j,\; i>j+k-1} f_N = p_k (\eta_j,
\eta_{j+1},\ldots ,\eta_{j+k-1})
\label{eq:trinv}
\end{equation}
for arbitrary $1 \leq j \leq N-k$. Consequently, the set of
$p_k(\eta_1, \ldots , \eta_k)$ functions satisfies the
following relations:
\begin{eqnarray}
p_k(\eta_1, \ldots , \eta_k)&=&\sum_{\eta_0} p_{k+1}(\eta_0,
\eta_1, \ldots , \eta_k)\ , \label{eq:cond1} \\
p_k(\eta_1, \ldots , \eta_k)&=&\sum_{\eta_{k+1}} p_{k+1}
(\eta_1, \ldots , \eta_k, \eta_{k+1})\ ,
\label{eq:cond2}
\end{eqnarray}
for $k>1$ and
\begin{equation}
\sum_{\eta_1} p_1(\eta_1)=1 \ .
\label{eq:norm}
\end{equation}
The probabilities of the $k$-point configurations are
completely described by introducing $2^{k-1}$ parameters
\cite{gut}. More precisely, the number of independent
variables may be less than $2^{k-1}$ when additional
symmetries (e.g. reflection, particle-hole) are taken into
consideration.

In $n$-point approximation the probability of a particle
configuration is expressed as:
\begin{equation}
f_N=p_n(\eta_1,\ldots ,\eta_n) \prod_{i=n+1}^{N}
P_1(\eta_i|\eta_{i-n+1},\ldots ,\eta_{i-1})
\label{eq:prod}
\end{equation}
where
\begin{equation}
P_1(\eta_i|\eta_{i-n+1}, \ldots ,\eta_i)= {p_n(\eta_{i-n+1},
\ldots ,\eta_i) \over p_{n-1}(\eta_{i-n+1},\ldots ,\eta_{i-1})}
\label{eq:condprod}
\end{equation}
describes the conditional probability of finding $\eta_i$
state at the $i$th site ($n<i\leq N$) if the cluster
configuration in the previous $n-1$ points is given. The
above expression is self-consistent in the sense that it
satisfies the condition of translation invariance [see
Eq.~(\ref{eq:trinv})] for $k \leq n$. To check it the
summation should be performed with respect to the first
and/or last site variables step by step using the
expressions (\ref{eq:cond1}) and (\ref{eq:cond2}).

Choosing the Boltzmann constant to be unit the entropy is
\begin{equation}
S=-\sum_{\{\eta_i\}}f_N \ln f_N \ ,
\label{eq:entropy}
\end{equation}
where the summation runs over all the possible configurations.
Substituting Eq.~(\ref{eq:prod}) into (\ref{eq:entropy}) one
obtains that
\begin{eqnarray}
S=&-&\sum_{\{\eta_i\}}f_N \ln p_n(\eta_1,\ldots
,\eta_n) \nonumber \\
 &-&\sum_{\{\eta_i\}}f_N \sum_{i=n+1}^{N} \ln P_1(\eta_i|\eta_{i-
n+1},\ldots ,\eta_{i-1}) \ .
\label{eq:entr}
\end{eqnarray}
According to Eq.~(\ref{eq:trinv}) this expression may be simplified
and in the thermodynamic limit the entropy per lattice sites is
\begin{eqnarray}
s&=&\lim_{N \to \infty} {S \over N} \label{eq:s} \\
&=&-\sum_{\eta_1 , \ldots ,\eta_n} p_n (\eta_1,\ldots ,
\eta_n) \ln \left( {p_n(\eta_1,\ldots ,\eta_n) \over p_{n-
1}(\eta_1,\ldots ,\eta_{n-1})} \right) \nonumber \ .
\label{eq:s1d}
\end{eqnarray}
As shown by Woodbury \cite{wood} this quantity is equivalent
to the conditional entropy introduced in information theory
\cite{sw}. According to Eqs. (\ref{eq:cond1}) and
(\ref{eq:cond2}) the above entropy may be written in the
following form:
\begin{equation}
s={\cal S}_n-{\cal S}_{n-1}
\label{eq:s1cvm}
\end{equation}
where
\begin{equation}
{\cal S}_n = - \sum_{\eta_1 , \ldots ,\eta_n} p_n(\eta_1,
\ldots , \eta_n) \ln {p_n(\eta_1, \ldots ,\eta_n)} \; .
\label{eq:SN}
\end{equation}
These expressions are very convenient for the CVM.

Using the techniques mentioned in the Introduction the
entropy of the one-dimensional system was previously derived
by several authors \cite{bethe,kiku,wood}. Here it is worth
mentioning that the pair ($n=2$) approximation reproduces
the exact solution for the one-dimensional Ising model with
nearest-neighbor interaction.

The present derivation is based on the fact that the
configuration probability given as product of conditional
probabilities is self-consistent, i.e. it satisfies the
Eq.~(\ref{eq:trinv}). The generalization of this approach
for higher dimensions is not trivial, approximations are
required as it will be shown in the following section.

\section{Square lattice}
\label{sec:square}

In order to explore the difficulties arising on a square
lattice we consider first the probability of a configuration
constructed on the analogy of the one-dimensional system
using square cluster configuration probabilities. In a
translation invariant system we introduce the function
$p_4(\eta_a, \eta_b, \eta_c, \eta_d)$ (clockwise labelling)
characteristic to the probability of any four-point
configurations on sites $a,b,c$ and $d$ forming a square
with nearest neighbor bond sides. From these quantities the
three-, two- and one-point configuration probabilities (on
the corresponding subclusters) can be derived on the analogy
of Eqs. (\ref{eq:cond1}) and (\ref{eq:cond2}). Henceforth we
restrict ourselves to systems having fourfold symmetry. In
this case the three- and two point configuration
probabilities are independent of the cluster orientation.

The system with $L \times M$ sites can be covered by $(L-1)
\times (M-1)$ overlapping squares. Following this covering
procedure the $f_N$ function is built up from $p_4(\eta_a,
\eta_b, \eta_c, \eta_d)$ functions. To visualize this
calculation a graphical representation of this product is
introduced as displayed in the subsequent figures. Here the
squares, triangles, solid lines and closed circles represent
$p_4$, $p_3$, $p_2$ and $p_1$ functions in the numerator
with site variable arguments corresponding to the positions.
If these quantities appear in the denominator then the above
symbols will be plotted by dashed lines or open circles. The
size of the closed (or opened) circles refer to the
exponents of $p_1$ functions which may differ from 1 (or -1)
in the examples investigated.

\begin{figure}
\centerline{\epsfxsize=8cm
            \epsfbox{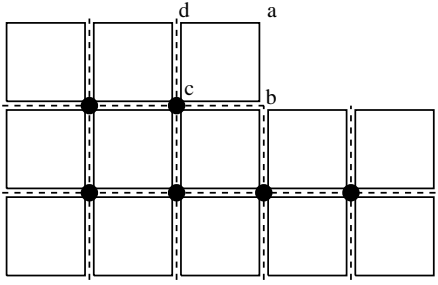}
            \vspace*{0.2mm}          }
\caption{Graphical representation of an intermediate stage
when constructing configuration probability as a product in
four-point approximation. Squares and closed circles
represent four-point and one-point cluster configuration
probabilities in the numerator with arguments corresponding
to their position. Dashed lines denote two-point
configuration probabilities in the denominator.}
\label{fig:sq4p}
\end{figure}

According to a direct (naive) way the addition of the site
variable $\eta_a$ to the product as shown in
Fig.~\ref{fig:sq4p} is performed via the following
conditional probability:
\begin{equation}
P_1(\eta_a | \eta_b, \eta_c, \eta_d)={p_4(\eta_a, \eta_b,
\eta_c, \eta_d) \over p_3(\eta_b, \eta_c, \eta_d)} \; .
\label{eq:naiv}
\end{equation}
Unfortunately, this construction results in a configuration
probability $f_N$ which is not self-consistent. This
difficulty may be circumvented if all the $p_3$ functions
appearing in the expressions are approximated as
\begin{eqnarray}
p_3(\eta_b, \eta_c, \eta_d)&=& \sum_{\eta_a} p_4(\eta_a,
\eta_b, \eta_c, \eta_d) \nonumber \\
&=&{p_2(\eta_b, \eta_c) p_2(\eta_c, \eta_d) \over
p_1(\eta_c)} \; .
\label{eq:sqapp4}
\end{eqnarray}
This situation is illustrated graphically in
Fig.~\ref{fig:sq4p}. Within the framework of this
approximation the self-consistency of the resultant $f_N$ is
easily checked because the summation defined in Eq.
(\ref{eq:sqapp4}) can be executed graphically too. As a
result the square transforms into an "angle" built up from
two $p_2$ and a $p_1$ functions. For example, the summation
over $\eta_a$ in Fig.~\ref{fig:sq4p} removes the $abcd$
square as well as the touched dashed lines and bullet. This
procedure may be repeated for all the $\eta_i$ variables
belonging to ``free'' site of a square until only the
desired square remains on the screen. As a consequence the
entropy per sites may be evaluated on the analogy of the
one-dimensional calculation. Neglecting boundary effects in
the thermodynamic limit it obeys the following form:
\begin{equation}
s={\cal S}_4-2 {\cal S}_2+{\cal S}_1
\label{eq:kw}
\end{equation}
where the ${\cal S}_n$ quantities are defined on the analogy
of Eq.~(\ref{eq:SN}) and the symmetries mentioned above are
taken into consideration. This formula is equivalent to
those derived by Kramers and Wannier \cite{kw}.

\begin{figure}
\centerline{\epsfxsize=8cm
            \epsfbox{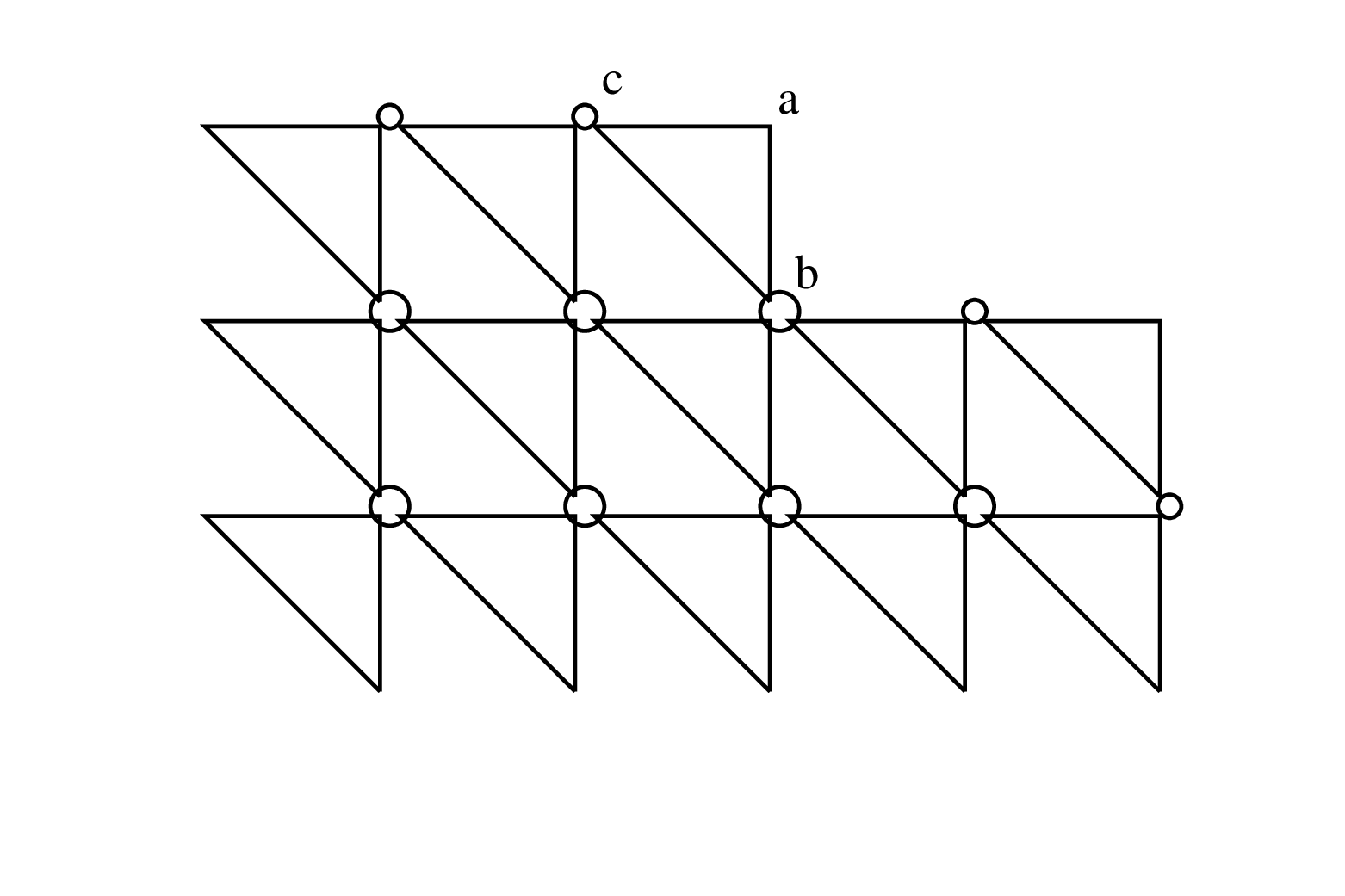}
            \vspace*{0.2mm}          }
\caption{In three-point approximation the configuration
probabilities are built up as a product of $p_3$ functions
(triangles) divided by $p_1^{\nu}$ terms where the value of
$\nu=1,2$ is denoted by the size of open circles.}
\label{fig:sq3p}
\end{figure}

The three-point (``angle'') approximation suggested by
Kikuchi \cite{kiku} can also be reproduced by the present
approach. In this case the lattice is covered with triangles
as shown in Fig.~\ref{fig:sq3p}.

The condition of self-consistency is fulfilled if the
two-point configuration
probabilities are approximated in the mathematical
manipulations as
\begin{eqnarray}
p_2^{\prime}(\eta_b, \eta_c)&=& \sum_{\eta_a} p_3(\eta_a,
\eta_b, \eta_c)= p_1(\eta_b) p_1(\eta_c) \; ,
\label{eq:sqapp3a} \\
p_2(\eta_a, \eta_c)&=& \sum_{\eta_b} p_3(\eta_a, \eta_b,
\eta_c)= p_1(\eta_a) p_1(\eta_c) \;
\label{eq:sqapp3b}
\end{eqnarray}
where we used the notation indicated in Fig.~\ref{fig:sq3p}
and similar formula is assumed for $p_2(\eta_a, \eta_b)$. On
the basis of Fig.~\ref{fig:sq3p} we can construct $f_N$ and
the entropy per lattice sites is
\begin{equation}
s={\cal S}_{3}-2{\cal S}_{1}
\label{eq:angle}
\end{equation}
Notice that in this approximation $f_N$ does not reflects
the fourfold symmetry assumed in the system in comparison
with those represented graphically in Fig.~\ref{fig:sq4p}.
This approximation takes into account all the pair
correlation via $p_3$, however, only a quarter of the
possible three-point correlations is handled.

\begin{figure}
\centerline{\epsfxsize=8cm
            \epsfbox{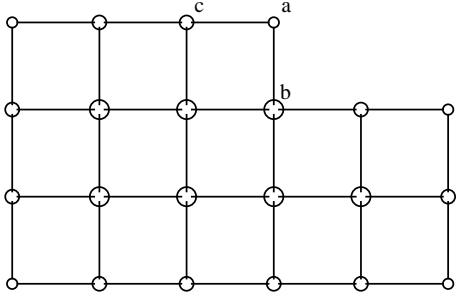}
            \vspace*{0.2mm}          }
\caption{Graphical representation of configuration
probability in pair approximation. Solid lines represent
two-point configuration probabilities in the numerator. The
sizes of open circles refer to the exponents (1, 2 or 3) of
$p_1$  functions in the denominator.}
\label{fig:sq2p}
\end{figure}

The reproduction of pair approximation is worth mentioning
for later convenience. One can image that in the above
three-point approximation the $p_3$ function is constructed
from $p_2$ functions as
\begin{equation}
p_3(\eta_a, \eta_b, \eta_c)={ p_2(\eta_a, \eta_b)
p_2(\eta_c, \eta_a) \over p_1(\eta_a)} \; .
\label{eq:pair1}
\end{equation}
To satisfy the condition of self-consistency one has to use
the following approximation:
\begin{equation}
\sum_{\eta_a} { p_2(\eta_a, \eta_b) p_2(\eta_c, \eta_a)
\over p_1(\eta_a)}=p_1(\eta_b) p_1(\eta_c) \; .
\label{eq:pair2}
\end{equation}
The graphical representation of the product construction is
illustrated in Fig.~\ref{fig:sq2p}. The entropy obtained
agrees with the result derived by Bethe \cite{bethe}.

We can choose larger clusters to increase the accuracy of
CVM. The above covering technique is required to have a
translation invariant $f_N$ constructed as a product of
conditional probabilities. The main problem is how to
construct the denominator of the conditional probability
characteristic to the probability of configurations on the
overlapping region. In general the following rule of thumb
seems to be useful for finding the approximations required
to satisfy the condition of self-consistency. The
configuration probability on the overlapping region should
be constructed from the largest subclusters of the present
clusters. This idea works well for $k \times k$-point
clusters. The graphical representation becomes confused if
one chooses large non-compact clusters. At the same time,
the numerical solution is also very complicate for large
clusters.

The generalization of the present approach to other (two- or
three-dimensional) Bravais lattices is straightforward. As
the simplest example for a non-Bravais lattice one can study
the chess-board like sublattice (antiferromagnetic) ordering
on the square lattice. In this situation the lattice is
divided into two ($A$ and $B$) interpenetrating sublattices
with different average occupations (or magnetizations). This
system can be described by introducing two types of
four-point configuration probabilities, namely
$p_4^A(\eta_a, \eta_b, \eta_c, \eta_d)$ and $p_4^B(\eta_a,
\eta_b, \eta_c, \eta_d)$ where the upper indices denotes the
sublattice to which site $a$ belongs. The entropy per sites
is derived by repeating the above calculation with using
alternately $p_4^A$ and $p_4^B$ ``squares'' during the
covering. The result is similar to those given by
Eq.~(\ref{eq:kw}), namely
\begin{equation}
s={1 \over 2}  \sum_{\alpha} \left[ {\cal S}^{\alpha}_{4}-
2{\cal S}^{\alpha}_{2}+{\cal S}^{\alpha}_{1} \right]
\label{eq:kwb}
\end{equation}
where $\alpha = A, B$ and fourfold symmetry is assumed.

\section{Tetrahedral sites in BCC lattice}
\label{sec:BCC}

\begin{figure}
\centerline{\epsfxsize=8cm
            \epsfbox{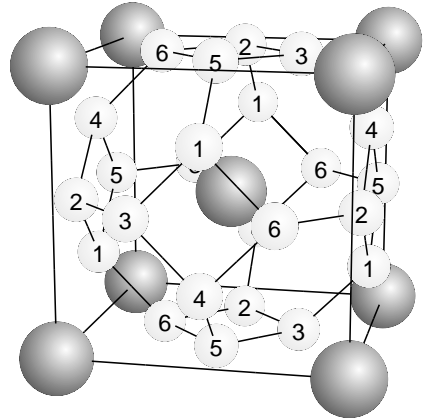}
            \vspace*{0.2mm}          }
\caption{Sublattice structure of tetrahedral interstitial
sites (small spheres) in BCC lattice (large spheres).}
\label{fig:bcc6}
\end{figure}

In the BCC lattice the interstitial atoms are positioned at
the tetrahedral sites exhibiting a non-Bravais lattice.
These sites can be divided into six (equivalent) sublattices
labelled by $\alpha=1, \ldots , 6$ (see
Fig.~\ref{fig:bcc6}).

In the six-sublattice mean-field approximation the
configuration probability is defined as a product of the
corresponding one-point configuration probabilities
$p_1^{\alpha}(\eta_i)$. This approximation is evidently
self-consistent and results in the following entropy
(henceforth normalized by the number of host lattice
points):
\begin{equation}
s=\sum_{\alpha} {\cal S}^{\alpha}_{1} \; .
\label{eq:bbc1}
\end{equation}
This approximation is used previously to demonstrate the
possibility of two subsequent phase transitions during the
ordering process in AgI \cite{AgI}. This method considers all the
phases conserving translation symmetry of the host lattice.
The cubic symmetry of these states is broken when the
sublattice occupations are different.

\begin{figure}
\centerline{\epsfxsize=8cm
            \epsfbox{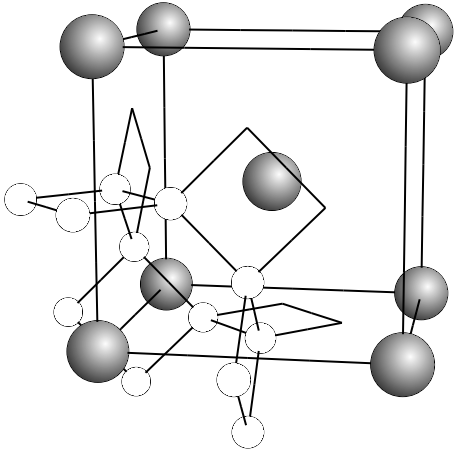}
            \vspace*{0.2mm}          }
\caption{Graphical representation of the conditional
probability when adding a cubic cell to the system. Squares
and open circles denote $p_4^{\beta}$ and $p_1^{\alpha}$
functions in the numerator and denominator.}
\label{fig:bcc4}
\end{figure}

In order to have a more accurate description of these states
we have to distinguish three different four-point clusters
positioned on the faces of a cubic cell. The probabilities
of the configurations in these clusters are denoted by
$p_4^{\beta}(\eta_a, \eta_b, \eta_c, \eta_d)$ where $\beta=
x, y, z$ refers to the face on which the square is
positioned. When the particle configuration is built up cube
by cube as a product of conditional probabilities then we
add 12 new site variables to the system. We choose these
variables to be located on the three connecting faces (see
Fig.~\ref{fig:bcc4}). The connections toward the touched
cubes are taken into account by the squares standing out the
cubic cell plotted in Fig.~\ref{fig:bcc4}. In fact, these
18-point clusters are used for covering. Now, however, the
probability of the configurations on these clusters are
constructed from $p_4^{\beta}$ and $p_1^{\alpha}$ functions
as represented graphically in the figure.

The open circles at the corners of the outstanding squares
denote that the configuration probability on the overlapping
region is approximated by a product of $p_1^{\alpha}$
functions in accordance with the rule of thumb mentioned
above. The $f_N$ function obtained is self-consistent if a
product $p_1^{\alpha}p_1^{\alpha^{\prime}}$ is substituted
for the two-point configuration probabilities appearing
during the summation process, for example:
\begin{equation}
\sum_{\eta_c,\eta_d}p_4^{\beta}(\eta_a, \eta_b, \eta_c,
\eta_d)=p_1^{\alpha}(\eta_a) p_1^{\alpha^{\prime}}(\eta_b)
\label{eq:bcc4}
\end{equation}
where $\alpha$ and $\alpha^{\prime}$ refers to the
corresponding sublattices. After some algebraic manipulation
the entropy is given as:
\begin{equation}
s= \sum_{\beta}  {\cal S}^{\beta}_4 -\sum_{\alpha}  {\cal
S}^{\alpha}_1 \; .
\label{eq:bccs4}
\end{equation}

One can choose different ways of covering for obtaining the
same result.  For example, first the $(x,y,0)$ sites are
covered by distinct horizontally oriented squares ($p_4^z$),
then we add all the vertical squares ($p_4^x$ and $p_4^y$)
to the system having two sites in the $(x,y,1/4)$ plane, and
so on.

Instead of suggesting other approximations for this lattice
structure, in the subsequent section we demonstrate how the
present method can be used for such a lattice exhibiting two
types of interstitial sites.

\section{Interstitial sites in FCC lattice}
\label{sec:FCC}

\begin{figure}
\centerline{\epsfxsize=8cm
            \epsfbox{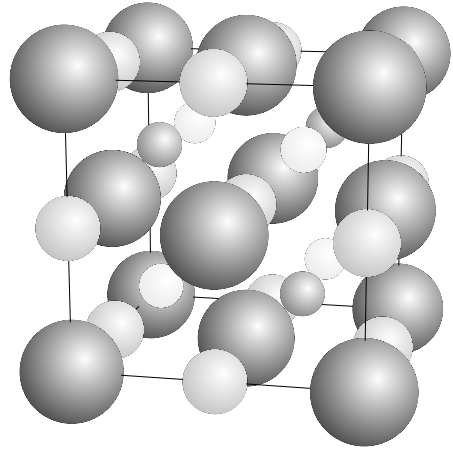}
            \vspace*{0.2mm}          }
\caption{The three-sublattice structure of tetrahedral
(small spheres with two gray scales) and octahedral (middle
spheres) interstitial sites in FCC lattice (large dark
spheres).}
\label{fig:fcc}
\end{figure}

In the FCC lattice there are one octahedral and two
tetrahedral interstitial sites per lattice points as
illustrated in Fig.~\ref{fig:fcc}. These sites can be
divided into three (FCC) sublattices. The three-sublattice
formalism allows us to study a rich variety of states
conserving the translation invariance of the host lattice.

\begin{figure}
\centerline{\epsfxsize=8cm
            \epsfbox{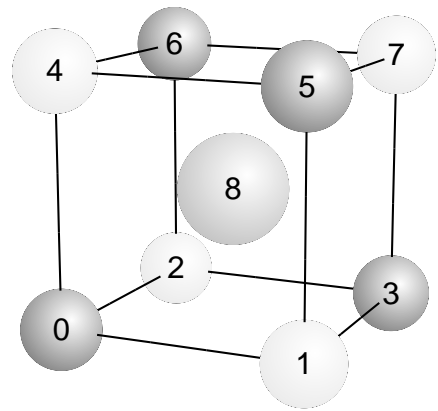}
            \vspace*{0.2mm}          }
\caption{Cubic (nine-point) cluster on interstitial sites in
FCC lattice. The figures on the spheres represent site
indices.}
\label{fig:fcc9}
\end{figure}

Notice that in this non-Bravais lattice the octahedral sites
are positioned at every second centers of the cubes
appearing in  NaCl-type structure of the tetrahedral sites.
This feature inspires the choice of simple cubic and
body-centered-cubic clusters (see Fig.~\ref{fig:fcc9}) for
the construction of $f_N$.

The condition of self-consistency is fulfilled if we use the
following approximations during summations:
\begin{eqnarray}
&&\sum_{\eta_7, \eta_8} p_9(\eta_0, \ldots , \eta_8)=
\sum_{\eta_7} p_8(\eta_0, \ldots ,\eta_7)= \label{eq:fcc9}
\\
&&{p_4(\eta_0, \eta_1, \eta_4, \eta_5) p_4(\eta_0, \eta_1,
\eta_2, \eta_3) p_4(\eta_0, \eta_2, \eta_4,
\eta_6)p_1(\eta_0) \over p_2(\eta_0, \eta_1) p_2(\eta_0,
\eta_2) p_2(\eta_0, \eta_4)} \nonumber
\end{eqnarray}
and similar expressions are assumed when instead of $\eta_7$
the sum runs over $\eta_i$ where $i \leq 6$. Here the site
indices of the arguments of $p_8$ agree with those shown in
Fig.~\ref{fig:fcc9} excepting 8. The same formulae will
appear in the denominators of the corresponding conditional
probabilities. Consequently, the entropy is
\begin{equation}
s={\cal S}_9+{\cal S}_8-6{\cal S}_4+6{\cal S}^{(T1-T2)}_2-
{\cal S}^{(T1)}_1-{\cal S}^{(T2)}_1
\label{eq:fcc9s}
\end{equation}
where the upper indices in the ${\cal S}_n$ functions
indicate sublattices to which the cluster sites belong if it
is necessery. In this expression the cubic symmetry is taken
into consideration.

\begin{figure}
\centerline{\epsfxsize=8cm
            \epsfbox{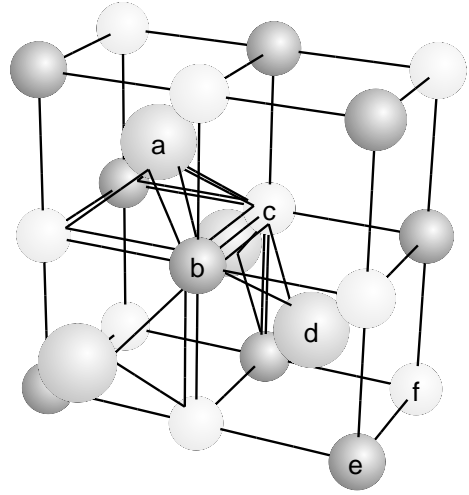}
            \vspace*{0.2mm}          }
\caption{Positions of $p_3$ functions connecting the triangle
$abc$ to its nearest neighbors having the same orientation
in case of triangle-triangle pair approximation.}
\label{fig:fcctri}
\end{figure}

Despite the symmetries of the above cubic clusters the
number of parameters will be large leading to difficulties
in the numerical calculations. The number of parameters is
drastically reduced if the cubic clusters (as well as the
whole system) are built up from smaller ones. In the
simplest case we can introduce a triangle cluster formed by
``nearest-neighbor'' sites belonging to different
sublattices (e.g. sites 0, 1 and 8 in Fig.~\ref{fig:fcc9}).
Suppose that $p_3(\eta_a,\eta_b,\eta_c)$ denotes the
probabilities of configurations on these clusters
independently of its orientation, where $\eta_a$, $\eta_b$
and $\eta_c$ belong to the octahedral and tetrahedral
sublattices, respectively. This choice is very convenient
because we have only seven parameters for the variation
technique. The configuration probabilities on the nine-point
cluster shown in Fig.~\ref{fig:fcc9} is expressed by these
quantities in such a way that the edges will be represented
with equal weight, that is
\begin{eqnarray}
p_9(\eta_0, \ldots , \eta_8)&=&{ p_3(\eta_8, \eta_0, \eta_1)
p_3(\eta_8, \eta_3, \eta_1) p_3(\eta_8, \eta_5, \eta_1)
\over p_2^2(\eta_8, \eta_2)} \nonumber \\
&\times&{p_3(\eta_8, \eta_3, \eta_7) p_3(\eta_8, \eta_3, \eta_2)
p_3(\eta_8, \eta_6, \eta_4)   \over
\tilde{p}_2^2(\eta_8, \eta_1) p_1(\eta_8) } \nonumber
\label{eq:fcc3}
\end{eqnarray}
where $p_2$ and $\tilde{p}_2$ are distinguished because
their second arguments belong to different tetrahedral
sublattices. The whole system may be covered by these
body-centered-cubic clusters overlapping each other at
single tetrahedral sites. In this approximation the entropy
is given as
\begin{eqnarray}
s=&6& {\cal S}_3-2 {\cal S}^{(O-T1)}_2-2 {\cal S}^{(O-T2)}_2
\nonumber \\
&-&{\cal S}^{(O)}_1 -3 {\cal S}^{(T1)}_1-3 {\cal S}^{(T2)}_1
\label{eq:fcc3s}
\end{eqnarray}
This triangle approximation, however, is not definite.
Another result may be obtained on the basis of Bethe's
method as follows. The elementary cell of this non-Bravais
lattice has three sites chosen to form a triangle cluster
defined above (see triangle $abc$ in Fig.~\ref{fig:fcctri}).

Now this triangle is considered as a single site variable
with $2^3$ possible values and the triangle-triangle pair
correlations are handled on the analogy of pair
approximation. For example, the configuration probability on
the triangle pair $abc$--$def$ is defined as a
$p_3p_3p_3/(p_2p_1)$ product. Due to the FCC symmetries each
triangle has twelve nearest neighbors. In this case, only
those triangle-triangle pairs are considered which are
connected to each other through a single triangle cluster as
illustrated in Fig.~\ref{fig:fcctri}.
As a result the entropy obeys the following form:
\begin{eqnarray}
s=&6& {\cal S}_3-2 {\cal S}^{(O-T1)}_2-2 {\cal S}^{(O-T2)}_2 -
{\cal S}^{(T1-T2)}_2 \nonumber \\
&-&{\cal S}^{(0)}_1-2{\cal S}^{(T1)}_1-2{\cal S}^{(T2)}_1
\label{eq:fcc33s}
\end{eqnarray}
which reproduces Eq.~(\ref{eq:fcc3s}) if
$p_1(\eta_b)p_1(\eta_c)$ is substituted for
$\hat{p}_2(\eta_b, \eta_c)$. Evidently, the approximations
required by the self-consistency are different in these
situations. The latter formula is used to check the mean-
field phase diagrams of a lattice-gas model suggested to
describe the ordering processes in alkali-fullerides
\cite{us}. The systematic comparison of the possible
approximations goes beyond the scope of the present work.

\section{Conclusions}
\label{sec:conc}

The derivation of entropy for cluster techniques is reformulated
by showing how to construct the configuration probabilities as a
product of conditional probabilities when building up the
system from overlapping clusters. In one-dimensional system
this product is obviously self-consistent. For higher
dimensional systems, however, some approximations are
required to fulfill the self-consistency which is relevant
when expressing the entropy per sites as a conditional
entropy introduced in information theory.

A graphical representation of the product is suggested to
make the calculations treatable. Using this technique we
have reproduced some traditional results derived on square
lattices. Due to its simplicity, this approach is easily
applicable to complicated lattices. Using this method
entropy expressions are derived on two non-Bravais lattices,
formed by the interstitial sites of BCC and FCC structures,
which may be useful for the investigation of many real
systems.

In fact, the present derivation of entropy is equivalent to
combinatorial calculations. However, it provides a better
understanding of the approximations applied. It is
emphasized that these methods assume merely translation
invariance despite some other techniques based on the
formalism of equilibrium statistical physics. As a
consequence the entropy expressions remain valid for those
non-equilibrium systems whose stationary states satisfy this
condition.

\acknowledgements

This research was supported by the Hungarian National
Research Fund (OTKA) under Grant No. T-4012.

\end{document}